\documentclass[aps,prb,twocolumn,superscriptaddress,a4paper,english]{revtex4}

\usepackage{amsfonts,epsfig}
\usepackage{amsmath}
\usepackage{epstopdf}
\usepackage{graphicx}
\usepackage{bm}
\usepackage{color}
\usepackage{amssymb}
\usepackage{ulem}
\usepackage{times}
\usepackage{dcolumn}
\usepackage{cases}
\usepackage{txfonts}
\usepackage[english]{babel}
\usepackage{epstopdf}
\usepackage{hyperref}
\usepackage{soul}

\DeclareGraphicsExtensions{.png,.eps}

\sethlcolor{yellow}

\begin{document}

\title{Controlling phase diagram of finite spin-$1/2$ chains by tuning boundary interactions}

\author{Shi-Ju Ran}
\email[Corresponding author. Email: ] {shi-ju.ran@icfo.eu}
\affiliation{Department of Physics, Capital Normal University, Beijing 100048, China}
\affiliation{ICFO-Institut de
	Ciencies Fotoniques, The Barcelona Institute of Science and
	Technology, 08860 Castelldefels (Barcelona), Spain}
\author{Cheng Peng}
\affiliation{School of Physical Sciences, University of Chinese Academy of Sciences, P. O. Box 4588, Beijing 100049, China}
\author{Gang Su}
\affiliation{School of Physical Sciences, University of Chinese Academy of Sciences, P. O. Box 4588, Beijing 100049, China}
\affiliation{Kavli Institute for Theoretical Sciences, University of Chinese Academy of Sciences, Beijing 100190, China}
\author{Maciej Lewenstein}
\affiliation{ICFO-Institut de Ciencies Fotoniques, The Barcelona Institute of Science and Technology, 08860 Castelldefels (Barcelona), Spain}
\affiliation{ICREA, Passeig Lluis Companys 23, 08010 Barcelona, Spain}

\begin{abstract}
  % For a quantum chain or wire, it is challenging to control or even alter the bulk properties or behaviors, for instance the response to external fields, by tuning solely the boundaries. This is expected, since the correlations usually decay exponentially, and do not transfer the physics at the boundaries to the bulk.
  Searching for simple models that possess non-trivial controlling properties is one of the central tasks in the field of quantum technologies. In this work, we construct a quantum spin-$1/2$ chain of finite size, termed as controllable spin wire (CSW), in which we have $\hat{S}^{z} \hat{S}^{z}$ (Ising) interactions with a transverse field in the bulk, and $\hat{S}^{x} \hat{S}^{z}$ and $\hat{S}^{z} \hat{S}^{z}$ couplings with a canted field on the boundaries. The Hamiltonians on the boundaries, dubbed as tuning Hamiltonians (TH's), bear the same form as the effective Hamiltonians emerging in the so-called ``quantum entanglement simulator'' that is originally proposed for mimicking infinite models. We show that tuning the TH's (parametrized by $\alpha$) can trigger non-trivial controlling of the bulk properties, including the degeneracy of energy/entanglement spectra, and the response to the magnetic field $h_{bulk}$ in the bulk. A universal point dubbed as $\alpha^s$ emerges. For $\alpha > \alpha^s$, the ground-state diagram versus $h_{bulk}$ consists of three ``phases'', which are Ne\'eL and polarized phases, and an emergent pseudo-magnet phase, distinguished by entanglement and magnetization. For $\alpha < \alpha^s$, the phase diagram changes completely, with no step-like behaviors to distinguish phases. Due to its controlling properties and simplicity, the CSW could potentially serve in future the experiments for developing quantum devices. % The CSW contains only nearest-neighboring spin-$1/2$ interactions, and could be realized in future experiments with atoms/ions coupled with artificial quantum circuits or dots.
  
  % the CSW surprisingly exhibits two different types of behaviors in response to the transverse field in the bulk. In one case, the staggered magnetization $M^s_z$ of the bulk exhibits a steep cliff, and the entanglement entropy $S$ shows a high peak. In the other case, no cliff or peak of $M^s_z$ and $S$ exists. We show that the behaviors of the CSW can be switched between these two types by solely tuning the TH's on the boundaries. 
\end{abstract}

% \pacs{75.78.Cd, 05.45.Pq, 85.35.Be}
\maketitle

\section{Introduction}

Quantum mechanics often leads to exotic behaviors and features that violate the common senses of the classical world. One of the central topics in modern science is to utilize the quantum features to build functional devices for the tasks that cannot be accessed, or efficiently solved by classical ones. Paradigm examples include simulating complicated quantum systems, or solving the NP-hard problems. Many efforts have been made towards this aim. One distinguished example concerns quantum simulators (cf. \cite{Qsimulator1,Qsimulator11,Qsimulator2,QsimulationRMP}, for the very recent achievements see \cite{Lukin17}), which are controllable quantum systems for efficiently mimicking the properties of other more complicated quantum systems. Another more universal proposal concerns quantum computers \cite{Qcomputer,QcomputerROPP}, which are expected to accelerate the exponentially expensive computations to be with only polynomial costs.

Rapid development in experimental techniques provides increasing possibility of and feasibility for realizing quantum technologies. For instance, cold/ultracold atoms in optical lattices or trapped ions \cite{Mbook,ColdAtom} allow to realize few-body models with designed interactions. Impressive results have been achieved using such methods to simulate many-body phenomena \cite{QPTQSblochRev} including ground-state phase transitions \cite{QPTQSbloch}, and/or dynamical processes \cite{SimuDynamicPRL} of large quantum lattice models.

However, there is still a long road to go for the practical implementations of quantum technologies. The proposals that can be applied in experiments or industry are rare. For example, the gate-based quantum computers \cite{QinfoBook} suffer high complexity, i.e., the demand of a large number of quantum gates. For Hamiltonian-based quantum annealers (e.g., D-wave \cite{Dwave}), the mainstream solution is to utilize the priori-known or constructed Hamiltonian such as Ising models or Kitaev model \cite{Kitaev}. A systematic ways of deriving new Hamiltonians for targeted quantum devices are strongly desired. Furthermore, most of the existing quantum technologies are based on few-body or few-level systems (see Refs. [\onlinecite{qc1, qc2, qc3, qc4}] to only name a few of the latest works). How to utilize the many-body features (such as quantum phase transitions, elementary excitations, and other collective phenomena) is still under hot debate (e.g., Refs. [\onlinecite{manybody1, manybody2}]).

% especially for utilizing the strongly-correlated systems: , and problems that can be solved by existing quantum devices in a more efficient manner than by the classical ones are also severely rare. Thus, searching for simple quantum models, and even more importantly, for systematic ways of designing interactions to realize non-trivial quantum devices is highly desired.

Recently, the ``quantum entanglement simulators'' (QES's) were proposed \cite{AOPHD}; these are few-body models that mimic optimally the ground states of the corresponding many-body systems of infinite size. A QES is formed by two parts: bulk and boundaries. The bulk is a finite-size super-cell of the infinite model to be simulated. The Hamiltonians on the boundaries give the optimal effective interactions between the boundary physical sites and entanglement-bath sites. These effective Hamiltonians are determined by the \textit{ab-initio} optimization principle (AOP) scheme \cite{AOP1D,tMPS}; they mimic optimally the entanglement between the finite bulk and the infinite environment in the targeted model.

In this work, we construct a one-dimensional (1D) spin-$1/2$ model of finite size, dubbed as controllable spin wire (CSW), where we have Ising interactions with a transverse field in the bulk, and the designed Hamiltonians (dubbed as the tuning Hamiltonians, TH's in short) for tuning on the boundaries. The TH's have the same form as the physical-bath Hamiltonians emerging in the QES, and are simply two-body nearest-neighboring spin-$1/2$ Hamiltonians with $\hat{S}^{x} \hat{S}^{z}$ and $\hat{S}^{z} \hat{S}^{z}$ interactions in a canted magnetic field [Eq. (\ref{eq-Hbath})].

We show that in the CSW, both the energy spectrum and ground-state entanglement spectrum, and the response of the bulk to the external field, can be qualitatively altered by tuning the TH's. A universal ``switch point'' (dubbed as $\alpha^s$) that addresses these qualitatively changes emerges. For $\alpha > \alpha^s$, the ground-state diagram versus $h_{bulk}$ consists of three phases, which are Ne\'eL and polarized phases, and an emergent pseudo-magnet phase, distinguished by entanglement and magnetization. In the ``pseudo magnet phase'', the energy gap closes exponentially with the system size, which is consistent with the results reported in Refs. [\onlinecite{BoundaryDef1,BoundaryDef2}], and the two-fold degeneracy of leading Schmidt numbers appears. For $\alpha < \alpha^s$, no step-like behaviors are observed to distinguish different phases. 

Thanks to the simplicity of the model, the Ising-type interactions of the CSW could be realized with, e.g., ultracold atoms or ions \cite{IsingExp1,IsingExp2,IsingExp3,IsingExp4,IsingExp5,IsingExp6}; the TH's (simply two-body) on the boundaries could be realized and tuned, for instance, by using super-lattices and ultracold atoms, ions in traps, quantum circuits or quantum dots coupled through photons with the bulk \cite{Mbook,Qcircuit1,Qcircuit2,Qcircuit3,Qcircuit4}.

\section{Hamiltonian of the controllable spin wire}
The Hamiltonian of the CSW with $N$ spin-$1/2$'s reads
\begin{eqnarray}
\hat{H} = \hat{H}_{Bulk} + \hat{H}_{L}(\alpha) + \hat{H}_{R}(\alpha),
\label{eq-HFB}
\end{eqnarray}
with $\alpha$ the control parameter. $\hat{H}_{Bulk}$ is the bulk Hamiltonian of the ($N-2$) spins in the middle of the CSW. The interactions are the nearest-neighbor Ising couplings in an uniform transverse field that reads
\begin{eqnarray}
  \hat{H}_{Bulk} = \sum_{n=2}^{N-2} \hat{S}^z_{n} \hat{S}^z_{n+1} - h_{Bulk} \sum_{n=2}^{N-1} \hat{S}^x_{n}.
  \label{eq-Hbulk}
\end{eqnarray}
The $\hat{S}^z \hat{S}^z$ coupling constant multiplied by the Planck constant $\hbar$ is set to be one, so that it defines the energy unit. Without losing generality, we take $N$ to be even.

\begin{figure}[tbp]
	\includegraphics[angle=0,width=1\linewidth]{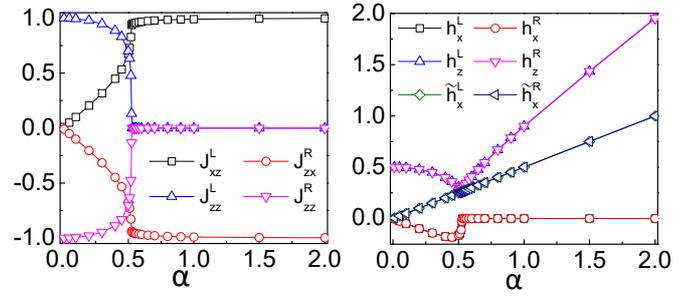}
	\caption{(Color online) The $\alpha$-dependence of the coupling constants (left) and magnetic fields (right) of the tuning Hamiltonians [Eq. (\ref{eq-Hbath})].}
	\label{fig-Hbath}
\end{figure}

$\hat{H}_{L}(\alpha)$ and $\hat{H}_{R}(\alpha)$ are the TH's, which give the interactions between the first two and the last two sites, respectively. These Hamiltonians contain $\hat{S}^{x} \hat{S}^{z}$ and $\hat{S}^{z} \hat{S}^{z}$ couplings in a canted magnetic field as
\begin{eqnarray}
\begin{aligned}
&\hat{H}_{L}(\alpha) = J^L_{xz} \hat{S}^x_1 \hat{S}^z_2 + J^L_{zz} \hat{S}^z_1 \hat{S}^z_2 - h^L_{x} \hat{S}^x_1 - h^L_{z} \hat{S}^z_1 - \tilde{h}^L_{x} \hat{S}^x_2,\\
&\hat{H}_{R}(\alpha) = J^R_{zx} \hat{S}^z_{N-1} \hat{S}^x_N + J^R_{zz} \hat{S}^z_{N-1} \hat{S}^z_{N} - h^R_{x} \hat{S}^x_{N} - h^R_{z} \hat{S}^z_{N} - \tilde{h}^R_{x} \hat{S}^x_{N-1}.
\end{aligned}
\label{eq-Hbath}
\end{eqnarray}
The coupling constants and magnetic fields depend on the controlling parameter $\alpha$, as shown in Fig. \ref{fig-Hbath}.

The TH's are parameterized by $\alpha$. Let us explain how we obtain the TH's and the $\alpha$-dependence of all the parameters of them. The idea is to borrow the physical-bath Hamiltonians emerging in the QES \cite{AOPHD}, which is originally for building a few-body model that optimally mimics the ground state of the infinite system and calculated by the AOP approach \cite{AOP1D,AOPHD}. Here, we take the TH's as the physical-bath Hamiltonian of the QES for the infinite transverse Ising model, whose Hamiltonian reads
\begin{eqnarray}
\hat{H}_{Inf} = \sum_{n} [\hat{S}^z_{n} \hat{S}^z_{n+1} - \frac{\alpha}{2} (\hat{S}^x_{n} + \hat{S}^x_{n+1})],
\label{eq-Hinf}
\end{eqnarray}
where the summation runs over the infinite chain. Note that $\alpha$ is a uniform transverse field in the $x$ direction but is now taken as the control parameter in the CSW. 

% A QES contains $\tilde{N}$ sites with $\tilde{N}$ a finite number, where the first and the last sites are called bath sites. The TH's on the boundaries of the QES are optimized, so that the reduced density matrix of the $(\tilde{N}-2)$ spins in the bulk of the QES optimally approximates the $\tilde{N}$-site reduced density matrix of the infinite model. Thus, the TH's are originally the effective Hamiltonians for simulating infinite models. 

The first and last sites in the CSW are corresponding to the bath sites of the QES. Here, we take the bath dimension as $\dim(bath)=2$, and use the TH's as physical Hamiltonians of spin-$1/2$'s. The detail of the algorithm can be found in the Appendix, as well as Refs. [\onlinecite{AOPHD,AOP1D,tMPS}]. After calculating the TH's \cite{SM}, we study the ground state of the Hamiltonian of the CSW in Eq. (\ref{eq-HFB}) by the exact diagonalization and finite-size density matrix renormalization group algorithm \cite{DMRG}.

% Since the TH's are from the QES, designed for simulating an infinite Ising chain, the transverse field $\alpha$ is a good choice as the control parameter of the TH's. 

%$\hat{H}_{L}$ and $\hat{H}_{R}$ are calculated from the infinite Hamiltonian $\hat{H}_{Inf}$ in Eq. (\ref{eq-Hinf}).

The $\alpha$-dependences of the coupling constants and magnetic fields in $\hat{H}_{L}$ and $\hat{H}_{R}$ are given in Fig. \ref{fig-Hbath}. Note that the Hamiltonian of the CSW is given by Eq. (\ref{eq-HFB}), and $\alpha$ only plays the role of a control parameter that has one-to-one correspondence with the coupling constants and fields of the TH's. In the following, we will tune the TH's by tuning $\alpha$.

Except the Ising interactions and the transverse field that originally exist in the infinite model, the $\hat{S}^x \hat{S}^z$ coupling and a vertical field emerge in the TH's. This could be interesting, because the $\hat{S}^x \hat{S}^z$ interaction is the stabilizer on the open boundaries of the cluster state, a highly entangled state that has been widely used in quantum information sciences \cite{ClusterState,ClusterState1}. % More relations with the cluster state are to be further explored.

Firstly, let us consider the magnetic fields $\tilde{h}_x^L$ and $\tilde{h}_x^R$ emerging on the second and the last second spins, respectively. Though these two terms are in $\hat{H}_{L}$ and $\hat{H}_{R}$, they in fact belong to $\hat{H}_{Bulk}$. To explicitly obey the translational invariance while computing the TH's, $\hat{H}_{Inf}$ is written as the summation of two-body terms as Eq. ({\ref{eq-Hinf}}). By taking a finite part from the infinite chain, one can see that the magnetic fields on both ends of this finite part is only half of the field on the other sites. Interestingly, this missing field automatically appears in $\hat{H}_{L}$ and $\hat{H}_{R}$, making the field uniform again. Our calculation confirms that $\tilde{h}^L_{x} = \tilde{h}^R_{x} = \alpha/2$. In the following, we remove these fields from $\hat{H}_{L}$ and $\hat{H}_{R}$, while ensuring that the field on all ($N-2$) sites in the bulk of the CSW is uniformly $h_{Bulk}$.

\begin{figure*}[tbp]
	\includegraphics[angle=0,width=1\linewidth]{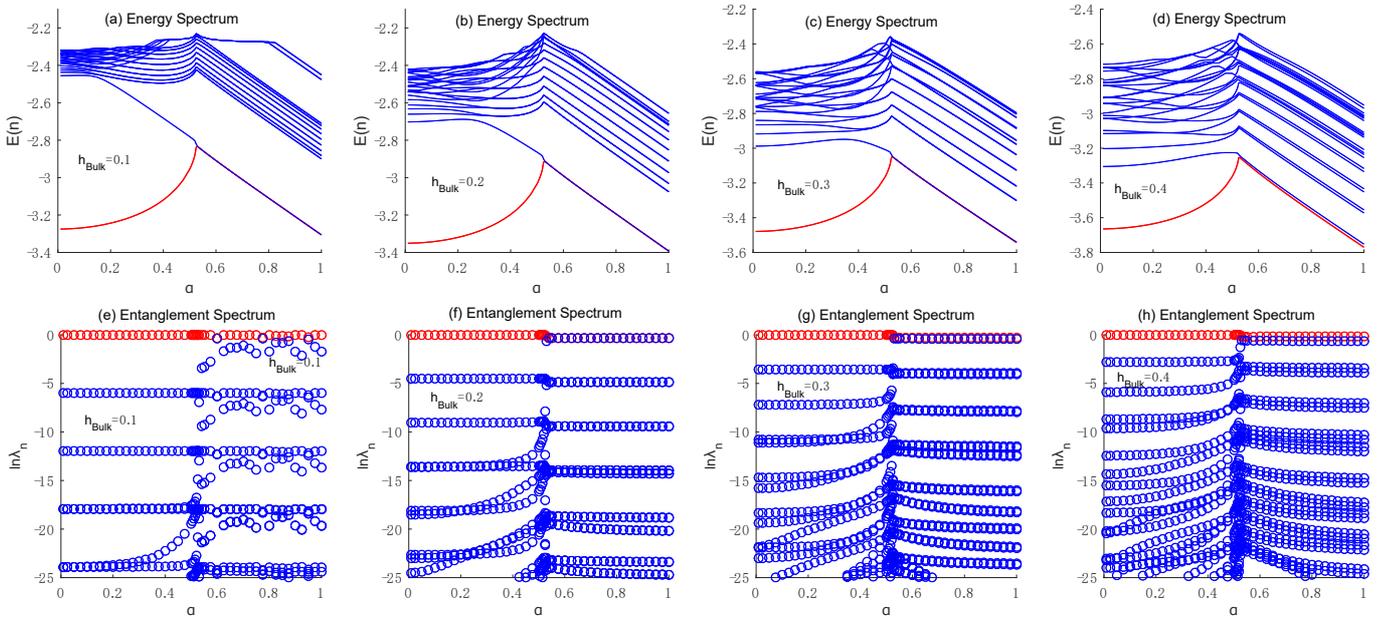}
	\caption{(Color online) The $\alpha$-dependence of (a)-(d) The energy spectrum and (e)-(h) the entanglement spectrum $\lambda$ of the ground state in the middle of the CSW with different $h_{Bulk}$. At $h_{Bulk}=0.1$, the ground-state energy has two-fold degeneracy (with a $\mathcal{O}(10^{-7})$ split) for $\alpha>0.5$, and the two leading Schmidt numbers show certain instability. At $h_{Bulk}=0.2$, the two-fold degeneracy appears for both the ground-state energy ($\mathcal{O}(10^{-5})$ split) and the entanglement spectrum ($\mathcal{O}(10^{-2})$ split). The split of the two dominant the energies or entanglement spectrum increases with $h_{Bulk}$ (e.g., $h_{Bulk}=0.3$ and $0.4$). In all cases, there is no degeneracy for $\alpha<0.5$. Here we take the length $N=12$.}
	\label{fig-ESL12}
\end{figure*}

From our results, we find that the number of parameters in the TH's can be reduced to five, which are
\begin{eqnarray}
\begin{aligned}
J^L_{xz} = -J^R_{zx} \doteq \bar{J}_{xz}, \ \ \ \  %\label{eq-Jbath1} \\
J^L_{zz} = - J^R_{zz} \doteq \bar{J}_{zz}, \label{eq-Jbath2}\\
h^L_{x} =  h^R_{x} \doteq \bar{h}_x, \ \ %\label{eq-Jbath3}\\
h^L_{z} = h^R_{z} \doteq \bar{h}_z, \ \  %\label{eq-Jbath4}\\
\tilde{h}^L_{x} = \tilde{h}^R_{x} \doteq \tilde{h}_{x}. \label{eq-J5}
\end{aligned}
\label{eq-Jbath5}
\end{eqnarray}
The reduction of the independent parameters in the TH's is due to the symmetry of the Hamiltonian. One can see that the spin-spin couplings show odd parity and the fields show even parity, when changing from the left boundary to the right. The reason should be that the couplings in the transverse Ising model is antiferromagnetic, and the field is uniform (not staggered). % Interestingly, changing the sign of $\bar{h}_x$ or $\bar{J}_{zz}$ does not affect the ground-state properties.

Let us take some limits of $\alpha$ to see how the values of the coupling constants in Eq. (\ref{eq-Jbath5}) change. As shown in Fig. \ref{fig-Hbath}, for $\alpha \to \infty$ we have $\bar{J}_{xz} \to 1$, $\bar{J}_{zz} \to 0$, $\bar{h}_x \to 0$, and $\bar{h}_z \to \alpha$. It means in this limit, the TH's become the $\hat{S}^x \hat{S}^z$ coupling in a vertical field (in the z direction). For $\alpha \to 0$, we have $\bar{J}_{xz} \to 0$, $\bar{J}_{zz} \to 1$, $\bar{h}_x \to 0$, and $\bar{h}_z \to 0.5$. The boundary interactions become the Ising coupling in a fixed vertical field. This leads exactly to the $N$-site classical Ising chain in a mean field.

%\begin{eqnarray}
%\begin{aligned}
%\bar{J}_{xz} \to 1,\ \ \
%\bar{J}_{zz} \to 0,\ \ \
%\bar{h}_x \to 0,\ \ \
%\bar{h}_z \to \alpha.
%\end{aligned}
%\label{eq-JlimitInf}
%\end{eqnarray}

%\begin{eqnarray}
%\begin{aligned}
%\bar{J}_{xz} \to 0,\ \ \
%\bar{J}_{zz} \to 1,\ \ \
%\bar{h}_x \to 0,\ \ \
%\bar{h}_z \to 0.5.
%\end{aligned}
%\label{eq-Jlimit0}
%\end{eqnarray}

Moreover, singular behaviors of the constants and fields are found near $\alpha=\alpha^s=0.5$. Interestingly, $\alpha=0.5$ is in fact the critical point of the infinite transverse Ising chain [Eq. (\ref{eq-Hinf})]. In the following, we will show that $\alpha^s=0.5$ is a ``\textit{switch point}'' where qualitative changes of the ground-state properties occurs.

%We discover that with the presence of the TH's, the ground state exhibits essentially different properties compared with a finite transverse Ising chain. 

\section{Controlling the energy/entanglement spectra}
Let us take $\alpha$ as the controlling parameter, and simulate how the CSW responds to $h_{Bulk}$. Fig. \ref{fig-ESL12} shows the $\alpha$-dependence of the energy spectrum $E(n)$ and the logarithmic ground-state entanglement spectrum $\ln \lambda_n$ with different $h_{Bulk}$. The entanglement is measured in the middle of the CSW. For $\alpha>0.5$ with small $h_{Bulk}$, e.g., $h_{Bulk}=0.1$, the excitation gap is around $\mathcal{O}(10^{-7})$. In this region, the entanglement spectrum show some instability. The reason could be that different super-positions of the degenerate states do not give a unique entanglement. Further discussions are presented in the supplementary material \cite{SM}.

At $h_{Bulk}=0.2$, the system still possesses two-fold degeneracy of the energy with a split around $\mathcal{O}(10^{-5})$. Meanwhile, the entanglement is stabilized and the two leading Schmidt numbers become degenerate with a split around $\mathcal{O}(10^{-2})$. As $h_{Bulk}$ becomes larger, e.g., $h_{Bulk}=0.3$ and $0.4$, the split increases, destroying the degeneracy of both the energy and entanglement spectra. In all cases, there is no degeneracy for $\alpha<0.5$. We dub $\alpha^s=0.5$ as the \textit{switch point}.

Though for $\alpha<\alpha^s$, the system gives a non-vanishing gap, thus has a unique ground state, we find that the ``degeneracy'' is hidden behind the sign of $\bar{J}_{zz}$. Since changing the sign of $\bar{J}_{zz}$ does not affect the ground-state properties, we calculate the fidelity to compare the ground states (Table \ref{tab-F}). We discover that the two Hamiltonians with different signs of $\bar{J}_{zz}$ have orthogonal ground states, i.e., the ground states are orthogonal to each other before and after changing the sign of $\bar{J}_{zz}$. For $\alpha>\alpha^s$ for comparison, the Hamiltonian has a vanishing gap, and changing the sign of $\bar{J}_{zz}$ does not make any differences. In the whole region, changing the sign of $\bar{h}_x$ does not affect anything. Our results suggest that the two orthogonal ground states can be controlled by tuning the sign of $\bar{J}_{zz}$ in the TH's. These degenerated states and the controlling properties can potentially serve to store and manipulate quantum information. Due to the gap beyond the degenerated states, such a process is expected to be stable against small noises. Surely, further numerical and experimental investigations are to be done to test the efficiency and robustness.

% We use $\hat{H}_{sign(\bar{h}_x)sign(\bar{J}_{zz})}$ to denote the Hamiltonians with different signs of $\bar{h}_x$ and $\bar{J}_{zz}$ (the absolute values of $\bar{h}_x$ and $\bar{J}_{zz}$, and other parameter are equal to each other).

\begin{table}[tbp]
	\caption{The fidelity $F=|\langle \phi' | \phi \rangle|$ between the ground states of two Hamiltonians that have different signs of $\bar{h}_x$ and $\bar{J}_{zz}$ in the TH's. We tale $N=18$, $\alpha=0.3$ and $h_{Bulk}=0.3$. The ground states of the Hamiltonians with opposite signs of $\bar{J}_{zz}$ have the same properties, but are orthogonal to each other ($F \sim \mathcal{O}(10^{-5})$).}
	\begin{center}
		\begin{tabular*}{8.5cm}{@{\extracolsep{\fill}}lcccc}
			\hline\hline % 插入横线，主要是美观上的考虑
			$\bar{h}_x / \bar{J}_{zz}$ & $+/+$ & $+/-$ & $-/+$ & $-/-$ \\ \hline
			$+/+$ & $1$ & $3.9(2) \times 10^{-5}$ & $0.97(5)$ & $4.3(8)\times 10^{-5}$  \\ \hline
			$+/-$ & $3.9(2) \times 10^{-5}$ & $1$ & $2.99 \times 10^{-5}$ & $0.97(5)$ \\ \hline
			$-/+$ & $0.97(5)$ & $2.99 \times 10^{-5}$ & $1$ & $3.9(2) \times 10^{-5}$ \\ \hline
			$-/-$ & $4.3(8)\times 10^{-5}$ & $0.97(5)$ & $3.9(2) \times 10^{-5}$ & $1$ \\
			\hline \hline
		\end{tabular*}
	\end{center}
	\label{tab-F}
\end{table}

\section{Controlling the ground-state phases}

\begin{figure}[tbp]
	\includegraphics[angle=0,width=1\linewidth]{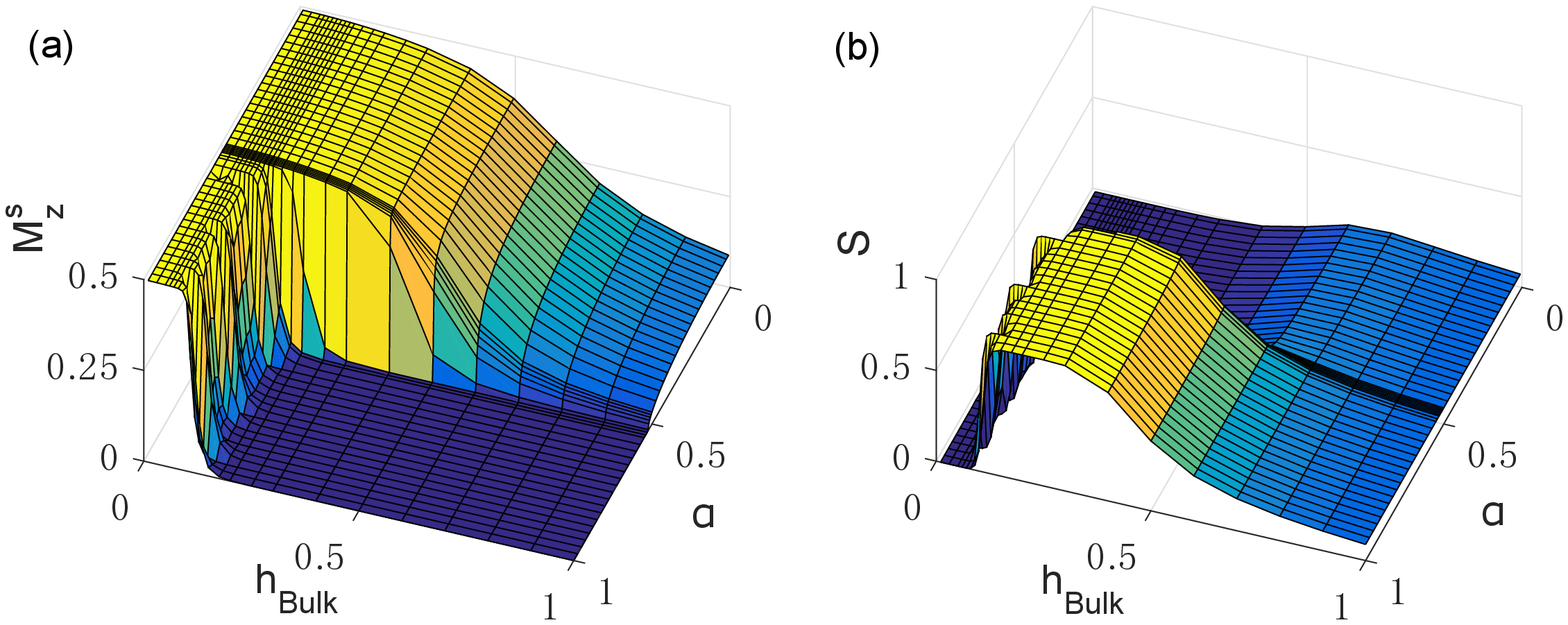}
	\includegraphics[angle=0,width=1\linewidth]{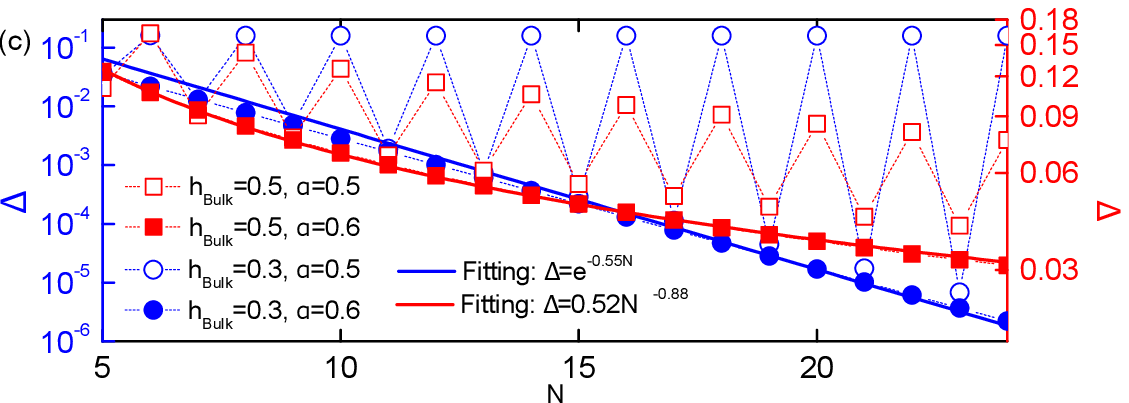}
	\caption{(Color online) The semi-log plot of the ground-state (a) staggered magnetization in the z direction $M^s_z$ and (b) the entanglement entropy $S$ with different values of the bulk transverse field $h^{Bulk}$ and the control parameter $\alpha$ of the TH's on the boundaries. We take $N=12$. (c) the size scaling of the excitation gap $\Delta$.}
	\label{fig-L8Phase}
\end{figure}

In Figs. \ref{fig-L8Phase} (a)-(b), we show the staggered magnetization in the z direction $M^s_z = \langle \phi|(\sum_{n=2,4,\cdots,N-2} \hat{S}^z - \sum_{n=3,5,\cdots,N-1} \hat{S}^z) |\phi \rangle / (N-2)$ and the entanglement entropy $S = -\sum \lambda_i^2 \ln \lambda_i^2$. These quantities show different kinds of responses to the transverse field $h_{Bulk}$ by taking different $\alpha$'s. For $\alpha>\alpha^s$, a step-like behavior occurs for $M^s_z$, which drops from $M^s_z \simeq 0.5$ to zero when $h_{Bulk}$ increases. We denote the position of the drop by $h^c_{Bulk}$. Similar singular behaviors are found for $S$ in this region, where $S$ reaches its maximum at $h=h^c_{Bulk}$. The peak of $S$ is from the two-fold degeneracy of the entanglement spectrum shown in Fig. \ref{fig-ESL12}, and thus can be the indicator of the degeneracy. For comparison, both $M^s_z$ and $S$ change smoothly with $h_{Bulk}$ for $\alpha<\alpha^s$. % Our work shows that the ground state can be switched between these landscapes by tuning the TH's on the boundaries.

Fig. \ref{fig-L8Phase} (c) shows the size scaling of the excitation gap $\Delta$. Then the $\alpha>\alpha^s$ region can be divided into three ``phases'' \cite{NotePhase}: conventional Ne\'el phase for $h_{Bulk}<h^c_{Bulk}$, conventional polarized phase for $h_{Bulk}>0.5$ (the critical point for the infinite Ising chain), and a ``pseudo-magnet'' (PM) phase for $h^c_{Bulk} <h_{Bulk}<0.5$. It is called ``magnet'' since $\Delta$ closes exponentially with $N$ \cite{BoundaryDef1,BoundaryDef2}; it is called ``pseudo'' because as $N \to \infty$, the PM phase vanishes when $h^c_{Bulk} \to 0.5$. The Ne\'el and PM phases both have exponentially vanishing gaps. The difference is that the PM phase has a robust entanglement arising from the two large leading Schmidt numbers. At $h_{Bulk}=0.5$, the excitation gap closes algebraically.
		
At $\alpha=\alpha^s$, the gap still closes for $h^c_{Bulk} <h_{Bulk}<0.5$, but only for odd $N$'s. For even $N$'s, the system is gapped with $\Delta \to \mathcal{O}(10^{-1})$. This possibly implies the valence bond(s) induced by the boundaries: a hanging spin appears to the edge when the length is odd. At $\alpha=\alpha^s$ and $h_{Bulk}=0.5$, the gap decrease algebraically for both the odd and even lengths.

% Note $h^c_{Bulk}$ is consistent with the singular point of the finite Ising chain without TH's.

%One can see that at $h_{Bulk}=h^c_{Bulk}$, the leading Schmidt numbers are stabilized and the difference between them reaches the minimum. Meanwhile, the two lowest energies remain degenerate at this point. As $h_{Bulk}$ continues increasing, both spectra start to split.

% For the ground-state uniform magnetization in the x direction $M^u_x = \langle \phi| \sum_{n=2}^{N-1}\hat{S}^z |\phi \rangle / (N-2)$, we find no singular behaviors in the whole $h_{Bulk}$-$\alpha$ region (see the supplementary material). With a fixed $h_{Bulk}$, $M^u_x$ changes very slowly when tuning $\alpha$. It means that the boundary interactions do not have much effects on $M^u_x$ in the bulk.

% {\R{Our results imply that at $h_{Bulk}=h^c_{Bulk}$ might give a topological ground state with non-trivial edge excitations (hasn't confirm yet).}}

% In the N\'eel region, $M^s_z$ decays slowly when $\alpha$ changes from $0$ to $\infty$. We have $M^s_z=0.5, 0.497, 0.484, 0.422$ at $\alpha=0,0.5,1,2$ and $h^{Bulk}=0$. According to Eq. (\ref{eq-JlimitInf}), we have that the vertical field $\tilde{h}_z$ increases linearly with other couplings approximately stay unchanged. These results suggest that the effects of simply increasing the field on the boundary are very small to the bulk.

We shall stress that with or without $\alpha$, these properties still exist and can be reached by setting the coupling constants and fields of the TH's according to the numerical results (Fig. \ref{fig-Hbath}). With $\alpha$, a well-defined switch point appears at $\alpha^s=0.5$. From Fig. \ref{fig-Hbath}, we have $\bar{J}_{xz} \simeq 1$, $\bar{J}_{zz} \simeq 0$, $\bar{h}_x \simeq 0$, and $\bar{h}_z \simeq 0.31$ at $\alpha=\alpha^s$. Comparing with the physics of the QES for the infinite quantum Ising chain in Eq. (\ref{eq-Hinf}), the switch point coincides with the critical point ($\alpha=0.5$) of the infinite model. Since the TH's optimally generate the entanglement bath of the infinite chain, our work implies that the control of the CSW should be triggered by the entanglement from the boundaries, which approximately mimics the entanglement from an infinite chain.

% For comparison, we implement similar numerical experiments on the finite transverse Ising chain without the TH's. The Hamiltonian simply reads $\hat{H}_{Ising} = \sum_{n=1}^{N-1} \hat{S}^z_{n} \hat{S}^z_{n+1} - h_{Bulk} \sum_{n=2}^{N-1} \hat{S}^x_{n} - \alpha (\hat{S}^x_{1} + \hat{S}^x_{N})$.
%\begin{eqnarray}
%\hat{H}_{Ising} = \sum_{n=1}^{N-1} \hat{S}^z_{n} \hat{S}^z_{n+1} - h_{Bulk} \sum_{n=2}^{N-1} \hat{S}^x_{n} - \alpha (\hat{S}^x_{1} + \hat{S}^x_{N}).
%\label{eq-FiniteIsing}
%\end{eqnarray}
% $h_{Bulk}$ is still the transverse field in the bulk, and $\alpha$ here is the transverse field on the boundary sites. As shown in Fig. \ref{fig-L8Phase} (c)-(d), no drastic changes occur for $M^s_z$ and $S$. The switch point is gone. It means that the response of the bulk to the transverse field on the boundaries becomes qualitatively weak. Furthermore, $\alpha$ does not alter the degeneracy of the energy spectrum or the entanglement spectrum \cite{SM}. Our results show that with the TH's, we are enabled to drastically alter the landscape of the bulk by only operating on the boundaries; without the TH's, one has to change the magnetic fields in the whole bulk to alter the properties of the system.

% The switch point is around $\alpha=0.5$, close to the genuine critical point of the infinite model, suggesting that the information of the criticality is encoded in the finite-size model by the boundary interactions.

More results are given in the supplementary material \cite{SM} to reveal the properties of the CSW, including the controlling effects of each single term in Eq. (\ref{eq-Hbath}), stability, size scaling, fidelity, and de-activation of the controlling. The differences and relations between our proposal and the boundary-driven phase transitions \cite{BoundaryDef1,BoundaryDef2} are further discussed. These could be useful information for the future applications of the CSW.

\section{Intuitive picture of controllable spin wire}

\begin{figure}[tbp]
	\includegraphics[angle=0,width=0.8\linewidth]{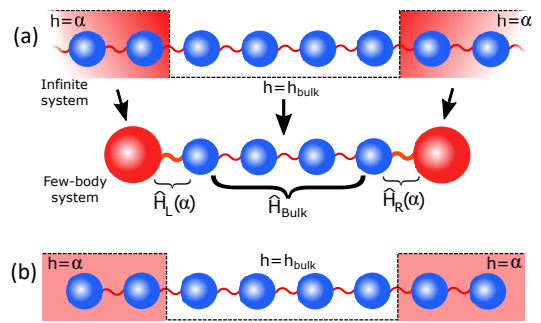}
	\caption{(Color online) (a) An intuitive picture of controllable spin wire. To control a finite bulk of an infinite chain without changing the interactions inside ($\hat{H}_{Bulk}$), one choice could be globally tuning the infinite rest of the chain. With the quantum entanglement simulator idea, the global tuning is approximately realized by the local tuning on the physical-bath Hamiltonians ($\hat{H}_{L(R)}(\alpha)$) on the boundaries of the bulk. (b) A normal finite-size model of size $(2+4+2)$, where a 4-site bulk is embedded in the middle of a 8-site chain.}
	\label{fig-CSWpic}
\end{figure}

We provide an intuitive picture [Fig. \ref{fig-CSWpic} (a)] to better explain the physics behind the CSW. Our aim is to control a finite chain by tuning the interactions on its boundaries. The proposal is to embed the system as a finite bulk in an infinite chain, and then the bulk (e.g., its response to a magnetic field) is controlled by tuning globally the infinite environment outside the bulk. However, this is extremely impractical by either numerics or experiments to realize directly this scheme. With the QES idea, the global tuning can be approximately realized by locally tuning the physical-bath Hamiltonians ($\hat{H}_{L(R)}(\alpha)$) solely on the boundaries (only one spin-$1/2$ on each end) of the bulk.

Furthermore, when the parameters inside and outside the bulk are tuned to favor different phases, a competition rises. This explains the universal switch point at $\alpha=0.5$, which is exactly the quantum phase transition point of the infinite model. Thanks to the generality of tensor network (see Refs. \onlinecite{AOP1D,AOPHD} or the first section of supplementary material), there are in principle no restrictions to the models for deriving the tuning Hamiltonians. We conjecture that the such controlling systems can be similarly constructed based on other models. Non-conventionally ordered systems (e.g., the integer spin chains \cite{Haldane1,Haldane2}) or systems in higher dimensions are to be inspected.

\begin{figure}[tbp]
	\includegraphics[angle=0,width=0.8\linewidth]{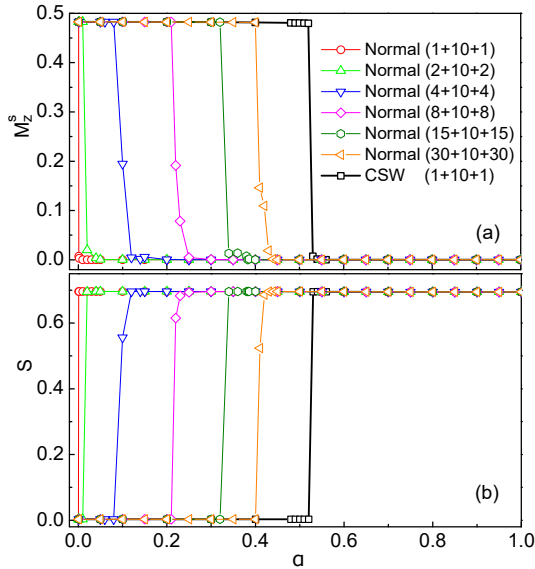}
	\caption{(Color online) The bulk magnetization $M_z^s$ and entanglement entropy $S$ of the CSW [see Fig. \ref{fig-CSWpic} (a)] and the quantum Ising chain without TH's [see Fig. \ref{fig-CSWpic} (b)]. Different sizes of the boundaries are calculated. When the boundaries increase, the control of the bulk $M_z^s$ and $S$ without TH's approaches to those of the CSW. We fix the length of the bulk to be 10 sites, and the magnetic field in the bulk to be $h_{Bulk} = 0.25$.}
	\label{fig-CSWvsChain}
\end{figure}

To further verify this intuitive picture, we calculate the normal quantum Ising chain with different length of the environment parts. An illustration of a 4-site bulk with 2-site boundaries on both ends is shown in Fig. \ref{fig-CSWpic} (b), which is dubbed as the normal $(2+4+2)$ chain. The magnetic field in the boundaries (dubbed still as $\alpha$) is tuned to be different from that inside the bulk (dubbed as $h_{Bulk}$).

Fig. \ref{fig-CSWvsChain} shows $M^s_z$ and $S$ of the normal $(K+10+K)$ chains by taking $K = 1$, $2$, $4$, $8$ and $15$. The bulk size is fixed to be $10$. The magnetic field $\alpha$ in the boundary parts varies from $0$ to $1$, and that in the bulk is fixed to $h_{Bulk}=0.25$. We find that as $K$ increases, the position where a step-like behavior occurs moves to $\alpha^s$ given by the CSW. This supports the intuitive picture.

\section{Conclusions and prospects}

We propose the controllable spin wire, with the quantum Ising interactions in the bulk and the QES-inspired interactions on the boundaries. By solely tuning the interactions on the boundaries, the bulk properties of the CSW (including the degeneracy, energy spectrum, ground-state entanglement, and magnetization) can be qualitatively altered. A universal ``switch point'' $\alpha^s=0.5$ is found to address the controlling effects. 

In the future, more feasible and tunable interactions that couple multiple CSW's into size-scalable networks are applicable. We expect to design and develop models with more complicated and universal controlling properties and realize them in experiments. It is also interesting to utilize the controlling effects to detect the bulk properties \cite{Impurity}.

% The interactions in the CSW are simple, and the system size can be moderately small, so that the CSW could be potentially used to build functional quantum devices with non-trivial controlling phenomena, using cold-atom or cold-ion platforms.

\section*{Acknowledgments} We acknowledge Ettore Vicari, Andrea Pelissetto, Leticia Tarruell, Ignacio Cirac, Tony J. G. Apollaro, Emanuele Tirrito, Angelo Piga, and Xi Chen for enlightening discussions. This work was supported by ERC AdG OSYRIS (ERC-2013-AdG Grant No. 339106), the Spanish MINECO grants FOQUS (FIS2013-46768-P), FISICATEAMO (FIS2016-79508-P), and ``Severo Ochoa'' Programme (SEV-2015-0522), Catalan AGAUR SGR 874, Fundaci\'o Cellex, EU FETPRO QUIC, EQuaM (FP7/2007-2013 Grant No. 323714), and CERCA Programme / Generalitat de Catalunya. S.J.R. acknowledges Fundaci\'o Catalunya - La Pedrera $\cdot$ Ignacio Cirac Program Chair. C.P. and G.S. were supported in part by the MOST of China (Grant No. 2013CB933401), the NSFC (Grant No. 14474279), and the Strategic Priority Research Program of the Chinese Academy of Sciences (Grant No. XDB07010100).


\begin{thebibliography}{99}
\bibitem{NotePhase} ``Phases'' is quoted because for finite-size models, there is no generic phase transition.
\end{thebibliography}
\end{document}